\documentclass[prd,showpacs,aps,prd,nofootinbib,floatfix,amsmath,amssymb]{revtex4}
\usepackage{graphicx}
\usepackage{slashed}
\usepackage{textcomp}
\begin{document}

\makeatletter
\newbox\slashbox \setbox\slashbox=\hbox{$/$}
\newbox\Slashbox \setbox\Slashbox=\hbox{\large$/$}
\def\pFMslash#1{\setbox\@tempboxa=\hbox{$#1$}
  \@tempdima=0.5\wd\slashbox \advance\@tempdima 0.5\wd\@tempboxa
  \copy\slashbox \kern-\@tempdima \box\@tempboxa}
\def\pFMSlash#1{\setbox\@tempboxa=\hbox{$#1$}
  \@tempdima=0.5\wd\Slashbox \advance\@tempdima 0.5\wd\@tempboxa
  \copy\Slashbox \kern-\@tempdima \box\@tempboxa}
\def\FMslash{\protect\pFMslash}
\def\FMSlash{\protect\pFMSlash}
\def\miss#1{\ifmmode{/\mkern-11mu #1}\else{${/\mkern-11mu #1}$}\fi}
\makeatother

\title{Gauge invariant electromagnetic properties of fermions induced by $CPT$ violation in the Standard Model Extension}
\author{A. Moyotl$^a$, H. Novales--S\' anchez$^b$, J. J. Toscano$^a$, and E. S. Tututi$^c$}
\address{$^a$Facultad de Ciencias F\'{\i}sico Matem\'aticas,
Benem\'erita Universidad Aut\'onoma de Puebla, Apartado Postal
1152, Puebla, Puebla, M\'exico.
\\ $^b$Divisi\'on de Ciencias e Ingenier\'ias, Universidad de Guanajuato Campus Le\'on, Loma del Bosque 103, Colonia Lomas del Campestre, 37150, Le\'on, Guanajuato, M\'exico.
\\$^c$Facultad de Ciencias F\'isico Matem\'aticas, Universidad Michoacana de San Nicol\'as de Hidalgo, Avenida Francisco J. M\'ujica S/N, 58060, Morelia, Michoac\'an, M\'exico.}
\begin{abstract}
Low--energy Lorentz--invariant quantities could receive contributions from a fundamental theory producing small Lorentz--violating effects. Within the Lorentz--violating extension of quantum electrodynamics, we investigate, perturbatively, the contributions to the one--loop $ff\gamma$ vertex from the $CPT$--violating axial coupling of a vector background field to fermions. We find that the resulting vertex function has a larger set of Lorentz structures than the one characterizing the usual, Lorentz invariant, parametrization of the $ff\gamma$ vertex. We prove gauge invariance of the resulting one--loop expression through a set of gauge invariant nonrenormalizable operators introducing new--physics effects at the first and second orders in Lorentz violation, and which generate tree--level contributions to the $ff\gamma$ vertex. Whereas loop contributions involving parameters that violate Lorentz invariance at the first order are $CPT$--odd, those arising at the second order are $CPT$--even, so that contributions to low--energy physics are restricted to emerge for the first time at the second order. In this context, we derive a contribution to anomalous magnetic moment of fermions, which we use to set a bound on Lorentz violation.
\end{abstract}


\maketitle

\section{Introduction}
\label{intro}
Lorentz symmetry is an essential ingredient in the formulation of most models aimed to describe nature at high energies. Currently, it is widely accepted that the successful Standard Model (SM) requires to be extended in order to comprise all high--energy--physics phenomena. Moreover, it has been shown that violation of Lorentz invariance may be engendered~\cite{sblss} in string theory and that it is intrinsic to noncommutative theories~\cite{lsvncft}. Even though violations of invariance under the discrete symmetries $C$, $P$, $T$, and most of their combinations have been customarily included as sources of new--physics effects in SM extensions, the interest in violations of Lorentz symmetry has attracted much interest only recently. 

During many years, experimental tests of Lorentz symmetry have been carried out, but the theoretical tool to systematically investigate possible violations of this cornerstone of physics was just recently formulated as an effective field theory description known as the Standard Model Extension (SME)~\cite{CK1}. Violations of Lorentz symmetry are incorporated by the SME through an infinite set of parameters that transform covariantly under observer Lorentz transformations, but which are invariant with respect to particle Lorentz transformations. It is the latter invariance which causes the partial breaking~\cite{CK2} of the particle Lorentz group. Since the SME coefficients can be generated by a fundamental theory including gravity, these quantities might be suppressed by powers of ratios of the electroweak scale to the Planck mass. Furthermore, up to today no experimental evidence of Lorentz--symmetry violations has been found, even with the aid of the most accurate experiments available nowadays, which indicates that the SME coefficients are tiny.

While the SME is a framework within which specific models~\cite{smsme1,smsme2,smsme3,smsme4} can be propounded, general model--independent investigations with emphasis on different sectors extending the SM can be performed. Many SME coefficients from all sectors have been, accordingly, bounded stringently~\cite{KR}. A subset of the SME that has attracted much attention, both theoretically~\cite{CK1,CK2,theosme3,theosme4,theosme5,theosme6,theosme7,theosme8,theosme9} and experimentally~\cite{expsme1,expsme2,expsme3,expsme4,expsme5,expsme6,expsme7,expsme8}, is the quantum electrodynamics (QED) sector. Notice, on the other hand, that effective field theory does not establish a limit for the number of terms that must be included in the SME, so that, in general, it contains renormalizable terms as well as nonrenormalizable ones~\cite{theosme8,nrsme2,nrsme3}. The set of all renormalizable terms receives the name of minimal SME.  In the present paper, we focus on the $CPT$--violating effects of the axial coupling of background field vectors $b_\nu$ to fermions~\cite{CK2}, present in the QED sector of the minimal SME, to calculate the corresponding one--loop contributions to the $ff\gamma$ vertex, with $f$ representing a fermion. We take the external fermions on shell, but leave the external photon off shell. We introduce $CPT$ violation perturbatively, for which we perform this calculation by taking into account only modifications of the fermionic propagator at the first order in $b_\nu$, while we utilize the ordinary Dirac equation. We calculate three Feynman diagrams with insertions of the $CPT$--violating two--point function produced by the renormalizable axial coupling of $b_\mu$ to fermions, added to the Dirac Lagrangian. Two of these diagrams involve only one insertion, but the third diagram has two, so that contributions at the second order in $b_\nu$ appear. Interesting works involving radiative corrections in the context of the SME exist~\cite{rcsme1,rcsme2,rcsme3,rcsme4,rcsme5,rcsme6,rcsme7,rcsme8,rcsme9}, and yet looking for Lorentz violation by means of loop calculations is fertile ground where clues of a fundamental description could manifest.

As the photon is a gauge field, the $ff\gamma$ vertex must be gauge invariant at any order of perturbation series. We have found that contraction of the one--loop Lorentz--violating $ff\gamma$ vertex function with the external momentum of the photon does not yield a simple Ward identity, even if the photon is taken on shell. To prove gauge invariance we build a set of gauge invariant nonrenormalizable terms that contain $b_\nu$--first-- and --second--order tree--level contributions to the $ff\gamma$ vertex. The general structure of such terms is
\begin{equation}
\label{geftt}
\frac{\eta^{(i)}_j}{m_f^{i-4}}\,(b_\nu)^k\,{\cal O}^{(i)}_j(f,A_\rho),
\end{equation}
where the $\eta^{(i)}_j$ are dimensionless coefficients and $m_f$ is the mass of the fermion under consideration. The whole factor $(b_\nu)^k\,{\cal O}^{(i)}_j(f,A_\rho)$ is a gauge invariant nonrenormalizable operator whose mass dimension is $i>4$, while $k$ is a power of the $b_\nu$ vector, so that it takes the values 1 or 2. The factor $(b_\nu)^k$ is in most cases Lorentz--contracted with ${\cal O}^{(i)}_j(f,A_\nu)$, which depends on the fermion and photon fields. These Lorentz contractions can be realized in several ways, which are not shown explicitly in Eq.~(\ref{geftt}). The index $j$ labels different terms with the same mass dimension. Nonrenormalizable terms at the first order in $b_\nu$ that generate Lorentz structures of the one--loop $ff\gamma$ vertex are $CPT$--odd, whereas $b_\nu$--second--order nonrenormalizable operators yielding such structures are invariant under $CPT$. After finding all contributing gauge invariant nonrenormalizable terms, whose mass dimensions range from five to nine, we extract the tree--level $ff\gamma$ couplings, then pass to vertex function space, identify gauge invariant terms, compare with the one--loop expression, and finally conclude that our result is gauge invariant.

The anomalous magnetic moment (AMM) is a physical quantity that has been calculated, in the SM, with remarkable accuracy for the cases of the electron~\cite{AHKNe,Krev} and the muon~\cite{AHKNm}. At present there are, in both cases, minuscule differences, $\Delta a_f$, between such theoretical predictions and the experimentally measured values, which makes it a place where suppressed new physics could manifest. Within the pure--photon sector of the SME, the AMM of the electron, emerged from the one-loop $ff\gamma$ vertex, was employed to constrain~\cite{rcsme7} isotropic Lorentz violation~\cite{theosme4,theosme5}. In the present paper, our calculation of the Lorentz--violating one--loop contributions to the $ff\gamma$ vertex, generated by the axial coupling of $b_\nu$ to fermions, yields no contributions to AMM at the first order in $b_\nu$, so we further look for it at the second order and find a clear contribution proportional to $b^2$, which allows us to constrain the magnitude of such background vector to be less than $10^{-9}$, for the case of the electron, and $10^{-5}$, for the case of the muon.

The contribution to the AMM that we derive in the present paper involves infrared divergences~\cite{BN}, which we extract and then omit in order set the bounds on Lorentz violation. The one--loop Lorentz--violating contributions to the $ff\gamma$ vertex that we calculate were also derived in Ref.~\cite{rcsme3}, where the authors computed the corresponding contributions to AMM and concluded that they are unphysical, since infrared divergences do not cancel at the level of cross section. In our opinion, the full theory might somehow cancel such divergences and render the results physically consistent, as it occurs, for instance, with the cancellation of ultraviolet divergences in effective theories via~\cite{efflagrangians} the $\overline{\rm MS}$ renormalization scheme~\footnote{Examples, in the context of effective field theory, in which cancellation of ultraviolet divergences is explicitly carried out in the $\overline{\rm MS}$ renormalization scheme can be found in Refs.~\cite{mscanc1,mscanc2}}. The determination of the presumable means that would eliminate these infrared divergences, though interesting, is beyond the scope of the present study.

We have organized the paper as follows. In Sec.~\ref{gqed}, we comment on general theoretical aspects and perform the calculation of the one--loop $ff\gamma$ vertex. We analyze gauge invariance of our results in Sec.~\ref{gieft}, where we provide a list of the gauge invariant nonrenormalizable terms that contribute to the $ff\gamma$ vertex at tree level and generate the Lorentz structures arisen in the one--loop calculation. In Sec.~\ref{dscn}, we extract the new--physics contribution to the AMM and set a bound on the Lorentz--violating effects associated to the electron and the muon. Finally, our conclusions are presented in Sec.~\ref{conc}.

\section{$CPT$--violating contributions to the one--loop $ff\gamma$ vertex}
\label{gqed}
Since there are physical phenomena not contained in the SM, incentives exist to look for a more complete description of nature. Effective field theory~\cite{efflagrangians} is a model--independent way to investigate effects of a presumable fundamental description at energies attainable by current accurate experiments. In general, effective theories are generated by integrating out heavy degrees of freedom in a physical description whose energy scale is large. The result is a hierarchical expansion, governed by low--energy symmetries and dynamical variables, that involves an infinite number of parameters. If the fundamental theory is not known since the onset, such parameters, which quantify the impact of the full theory at low energies, can be estimated or bounded by comparison with experimental data. The SME is a general framework, built within effective field theory, which, as such, offers the possibility of studying, at energies within the reach of current experiments, the effect of a high--energy formulation in which spontaneous breaking of Lorentz symmetry occurs~\cite{K}.

The QED sector of the minimal SME incorporates Lorentz--symmetry violation through the pure--photon sector and the fermionic sector as well. The full set of SME coefficients can be classified~\cite{CK1} according to whether their corresponding terms are even or odd under the discrete transformation $CPT$. In the present paper, we are particularly interested in $CPT$--odd effects coupling axially in the renormalizable fermionic sector of QED, and which are parametrized by four coefficients distributed in one four--vector. The QED Lagrangian including, exclusively, such $CPT$--violating axial coupling is given by~\cite{CK1,CK2}
\begin{equation}
{\cal L}_{\rm QED}^{CPT-{\rm odd}}=\bar{\psi}\left( i\gamma^\mu D_\mu+\gamma^\mu b_\mu\gamma^5-m_f \right)\psi-\frac{1}{4}F^{\mu\nu}F_{\mu\nu}-\frac{1}{2}\,\partial^\mu A_\mu\,\partial^\nu A_\nu,
\label{rcptvb}
\end{equation}
where $D_\mu=\partial_\mu+ieA_\mu$ is the gauge covariant derivative, $F^{\mu\nu}=\partial^\mu A^\nu-\partial^\nu A^\mu$ is the field strength associated to the photon gauge field $A^\mu$, $\psi$ is a fermionic field whose charge is that of the electron, and $b_\nu$ is the background field responsible for violation of Lorentz symmetry. An important observation is that, in general, the $CPT$--violating vector $b_\nu$ is not~\cite{CK1} universal for all fermions, but a different vector corresponds to each fermion. Note that we have added in Eq.~(\ref{rcptvb}) a gauge--fixing term that corresponds to the Feynman--'t Hooft gauge.


\begin{figure}[!ht]
\center
\includegraphics[width=5.5cm]{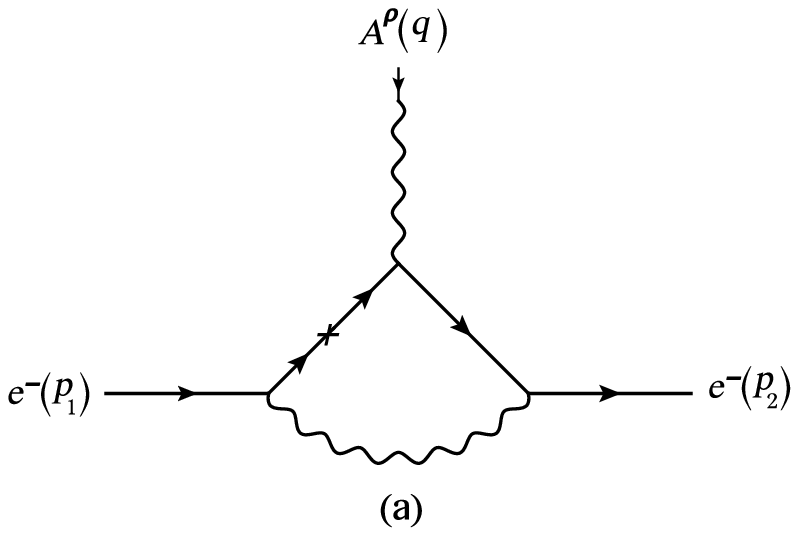}
\hspace{0.5cm}
\includegraphics[width=5.5cm]{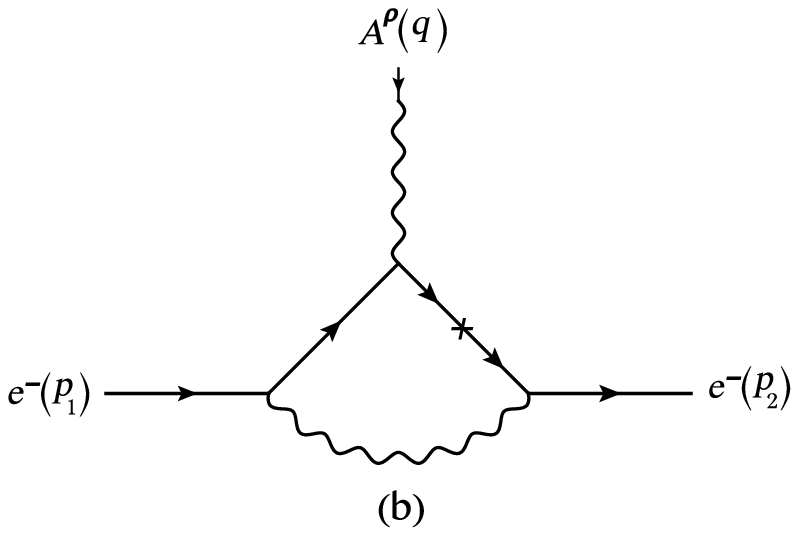}
\hspace{0.5cm}
\includegraphics[width=5.5cm]{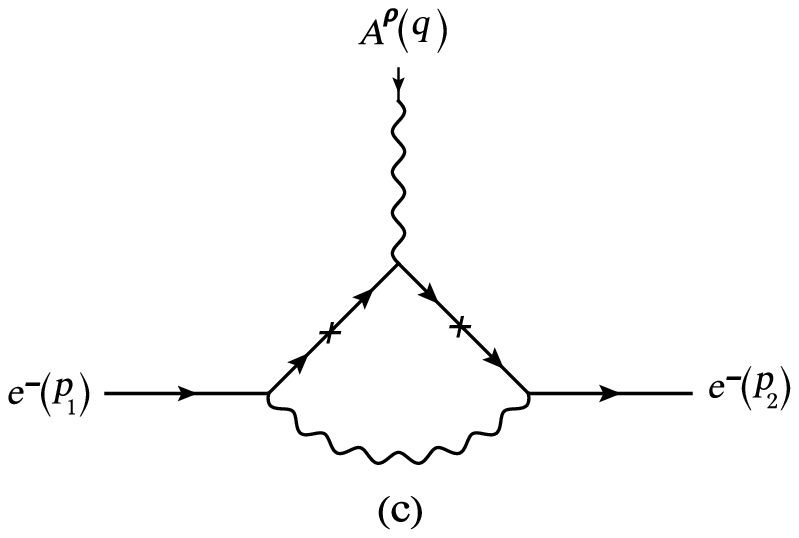}
\caption{\label{eegamma} One--loop diagram contributing to the $ff\gamma$ vertex.}
\end{figure}
The presence of the $b_\nu$ background vector field induces modifications of the free fermionic propagator. The loop calculation that we pursue in the present work is carried out perturbatively, so, instead of using the exact~\cite{rcsme1,rcsme2} version of such modified propagator, we obtain, from Eq.~(\ref{rcptvb}), the two--point vertex function produced by the $CPT$--violating axial coupling and insert it in internal fermionic lines of the one--loop diagram contributing to $ff\gamma$. This is equivalent to assume in the exact modified fermionic propagator that the vector $b_\nu$ is tiny, and then take an approximation at the first order in $b_\nu$~\cite{rcsme2}. Available bounds on Lorentz--violating fields $b_\nu$ exist~\cite{KR}, and such bounds establish that the components of the corresponding vectors are rather small. So, this perturbative approach is well founded. The three diagrams that we calculate are provided in Fig.~(\ref{eegamma}), where the $CPT$--violating insertions are represented by small crosses on the internal fermionic lines. The sum of these diagrams can be written as a sum of two terms:
\begin{equation}
ie\,\Gamma_\mu^{ff\gamma}=ie\left( \Gamma_\mu^b+\Gamma_\mu^{b^2} \right).
\end{equation}
The first term of this expression, which is produced by diagrams (a) and (b) of Fig.(\ref{eegamma}), incorporates all $b_\nu$--first--order contributions to the $ff\gamma$ vertex at the one--loop level, and is given as the sum of the two loop integrals
\begin{eqnarray}
\Gamma_\mu^b&=&ie^2\int\frac{d^4k}{(2\pi)^4}\,\frac{\gamma^\nu(\slashed{k}+\slashed{p}_2+m_f)\gamma_\mu(\slashed{k}+\slashed{p}_1+m_f)\slashed{b}\gamma^5(\slashed{k}+\slashed{p}_1+m_f)\gamma_\nu}{\big[ (k+p_1)^2-m_f^2 \big]^2\big[ (k+p_2)^2-m_f^2 \big]k^2}
\nonumber \\ &&
+ie^2\int\frac{d^4k}{(2\pi)^4}\,\frac{\gamma^\nu(\slashed{k}+\slashed{p}_2+m_f)\slashed{b}\gamma^5(\slashed{k}+\slashed{p}_2+m_f)\gamma_\mu(\slashed{k}+\slashed{p}_1+m_f)\gamma_\nu}{\big[ (k+p_1)^2-m_f^2 \big]\big[ (k+p_2)^2-m_f^2 \big]^2k^2}.
\end{eqnarray}
The superficial degrees of divergence of the integrals in $\Gamma^b_\mu$ indicate that no ultraviolet divergences can arise from them. All $b_\nu$--second--order contributions, arisen from diagram ${\rm ( c )}$ of Fig.~(\ref{eegamma}), are contained in $\Gamma_\mu^{b^2}$, whose explicit expression is
\begin{eqnarray}
\Gamma_\mu^{b^2}&=&ie^2\int\frac{d^4k}{(2\pi)^4}\,\frac{1}{\big[ (k+p_1)^2-m_f^2 \big]^2\big[ (k+p_2)^2-m_f^2 \big]^2k^2}
\gamma^\nu(\slashed{k}+\slashed{p}_2+m_f)\slashed{b}\gamma^5
\nonumber \\&&
\times
(\slashed{k}+\slashed{p}_2+m_f)\gamma_\mu(\slashed{k}+\slashed{p}_1+m_f)\slashed{b}\gamma^5(\slashed{k}+\slashed{p}_1+m_f)\gamma_\nu.
\end{eqnarray}
Again, the superficial degree of divergence ensures that this loop integral is ultraviolet--convergent. Note that this last term contains two $\gamma^5$ matrices, which cancel each other, so that, since the beginning, it is not expected to produce contributions to electric dipole moment of fermions. As we take the external fermions on shell, the Dirac equation may be employed in the calculation. Notice that the presence of the $CPT$--violating vector field $b_\nu$ modifies~\cite{CK2} this equation. However, within our perturbative treatment, we use the ordinary Dirac equation through the whole process.

We first calculate the Lorentz--violating $b_\nu$--first--order contributions. Using the Feynman--parameters technique, we express the term $\Gamma^b_\mu$ as
\begin{eqnarray}
\Gamma^b_\mu&=&\frac{1}{m_f}i\sigma_{\mu\nu}b^\nu\gamma^5\,F_1^b(q^2)+
\frac{1}{m_f^2}b\cdot(p_1+p_2)\,\gamma_\mu\gamma^5\,F^b_2(q^2)+\frac{1}{m_f^2}(p_{1\mu}+p_{2\mu})\,\slashed{b}\gamma^5\,F^b_3(q^2)
\nonumber \\ &&
+\frac{1}{m_f^3}i(b\cdot q)\,\sigma_{\mu\nu}q^\nu\gamma^5\,F_4^b(q^2)+\frac{1}{m_f^3}\,b\cdot(p_1+p_2)\gamma^5q_\mu\,F_5^b(q^2),
\label{order1r1}
\end{eqnarray}
where the $F^b_j(q^2)$ are dimensionless factors that involve parametric integrals. We have annexed the precise expressions of these factors in Appendix~\ref{fopi}. The usual Lorentz--invariant parametrization of the $ff\gamma$ vertex~\cite{ffapar}, with the external fermions on shell, is made up of four terms, which correspond to the vectorial and axial currents, the magnetic dipole moment, and the electric dipole moment. Nevertheless, none of the Lorentz structures constituting Eq.~(\ref{order1r1}) coincide with terms of such well--known result. For instance, at the first glance the shape of the fourth term of $\Gamma_\mu^b$ resembles the Lorentz structure that defines the electric dipole form factor. If one naively tries to extract such form factor from this term, the resulting expression is an imaginary number. We further comment on this issue in the next section.

The second--order contributions of the $CPT$--violating vector $b_\nu$, all of them located in the $\Gamma_\mu^{b^2}$ term, has a larger number of different Lorentz structures than $\Gamma_\mu^{b}$. As before, we use the Feynman--parameters technique to calculate this term, which we find to be
\begin{eqnarray}
\Gamma_\mu^{b^2}&=&\frac{1}{m_f^2}\,b^2\gamma_\mu \,F_1^{b^2}(q^2)+\frac{1}{m_f^2}\,b_\mu\slashed{b}\,F^{b^2}_2(q^2)+\frac{1}{m_f^3}\,ib^2\sigma_{\mu\nu}q^\nu\,F^{b^2}_3(q^2)
\nonumber \\&&
+\frac{1}{m_f^3}\,b_\mu b\cdot(p_1+p_2)\,F^{b^2}_4(q^2)
+\frac{1}{m_f^3}\,i(b\cdot q)\,\sigma_{\mu\nu}b^\nu\,F^{b^2}_5(q^2)
+\frac{1}{m_f^4}\,b^2q^2\gamma_\mu\,F^{b^2}_6(q^2)
\nonumber \\&&
+\frac{1}{m_f^4}\,\gamma_\mu\left[ (b\cdot p_1)^2+(b\cdot p_2)^2 \right]\,F^{b^2}_7(q^2)
+\frac{1}{m_f^4}\,\gamma_\mu(b\cdot p_1)(b\cdot p_2)\,F^{b^2}_8(q^2)
\nonumber \\&&
+\frac{1}{m_f^4}\,(b\cdot q)\,\slashed{b}q_\mu\,F^{b^2}_9(q^2)
+\frac{1}{m_f^4}\,\slashed{b}\left[ p_{1\mu}(b\cdot p_1)+p_{2\mu}(b\cdot p_2) \right]\,F^{b^2}_{10}(q^2)
\nonumber \\&&
+\frac{1}{m_f^5}\,i(b\cdot p_1)(b\cdot p_2)\,\sigma_{\mu\nu}q^\nu\,F^{b^2}_{11}(q^2)
+\frac{1}{m_f^5}\,\left[ p_{1\mu}(b\cdot p_1)^2+p_{2\mu}(b\cdot p_2)^2 \right]\,F^{b^2}_{12}(q^2)
\nonumber \\ &&
+\frac{1}{m_f^5}\,\left[ p_{1\mu}(b\cdot p_2)^2+p_{2\mu}(b\cdot p_1)^2 \right]\,F^{b^2}_{13}(q^2),
\label{sovf}
\end{eqnarray}
where the $F_j^{b^2}(q^2)$ factors, exhibited explicitly in Appendix~\ref{sopi}, are dimensionless and are defined in terms of parametric integrals. Notice that the third term of this vertex function produces a contribution, proportional to $b^2$, to the AMM, which we discuss in more detail in Sec.~\ref{dscn}. There is also a ultraviolet--finite contribution to the vector current in the first term and a contribution to anapole moment in the sixth term. There are other terms that, according to their corresponding Lorentz structures, seem to introduce further contributions to the AMM and the vectorial current. However, they do not, as we discuss in the next section. All other Lorentz structures are new.

Up to this point, we have derived all first-- and second--order $CPT$--violating contributions associated to the modifications of the free fermionic propagator that were induced by the vector field $b_\nu$. Recall that, in obtaining this result, the photon was assumed to be off shell. Before analyzing the contributions to AMM arisen at the second order in $b_\nu$, we discuss on gauge invariance of our result. One can straightforwardly check that, even taking the photon on shell, contracting either $\Gamma^b_\mu$ or $\Gamma^{b^2}_\mu$ with the momentum $q$ of the external photon does not produce a simple Ward identity, that is, $q^\mu\Gamma_\mu^{b}\neq0$ and $q^\mu\Gamma_\mu^{b^2}\neq0$. We carry out, in the next section, the proof of gauge invariance in a different fashion that relies on the gauge invariant structure of nonrenormalizable terms in effective field theory. Notice that all terms in $\Gamma_\mu^b$ and $\Gamma_\mu^{b^2}$ are divided by a power of the fermionic mass, which resembles the nonrenormalizable terms in an effective Lagrangian. Indeed, such inverse powers of fermion mass indicate which is the mass dimension of each nonrenormalizable term of the effective Lagrangian containing the $CPT$--violating four vector $b_\nu$ and contributing, at tree level, to the $ff\gamma$ vertex through the Lorentz structures that are present in the one--loop expression. In the next section, we show such contributing nonrenormalizable operators.

\section{Gauge invariance}
\label{gieft}
All terms of any effective Lagrangian engendered, after integrating out heavy degrees of freedom, by a high--energy depiction are governed by low--energy symmetries. If the precise form of a presumable high--energy description is unknown, one can always construct an effective Lagrangian by using low--energy dynamic variables and imposing low--energy symmetries on every single term. Most effective Lagrangians that are made up of SM fields are constructed under the assumption that Lorentz symmetry and gauge symmetry associated to the ${\rm SU}(3)_{\rm C}\times{\rm SU}(2)_{\rm L}\times{\rm U}(1)_Y$ group are fulfilled. It has been a common practice to allow violation of discrete symmetries related to $C$, $P$, and $T$ in effective Lagrangians, although maintaining $CPT$ invariance. In the case of the SME, the requirement of Lorentz symmetry is relaxed by allowing violations associated with the particle Lorentz group, which in some cases also involves~\cite{G} violations of the $CPT$ discrete symmetry. In spite of this, gauge invariance is still assumed to be a symmetry of the effective field theory description. One is then able to construct nonrenormalizable operators that violate Lorentz--symmetry, but which are gauge invariant. In this section, we use this fact and provide all gauge invariant nonrenormalizable operators that contribute, at tree level, to the $ff\gamma$ vertex function that we just calculated.

In what follows we use the definitions
\begin{eqnarray}
\bar{\psi}\stackrel{\leftrightarrow}{\nabla}\Gamma\psi&\equiv&\bar{\psi}\Gamma(\nabla\psi)-(\overline{\nabla\psi})\Gamma\psi,
\\ \nonumber \\
\bar{\psi}\stackrel{\Leftrightarrow}{\nabla}\Gamma\psi&\equiv&\bar{\psi}\Gamma(\nabla\psi)+(\overline{\nabla\psi})\Gamma\psi,
\end{eqnarray}
with $\nabla$ representing any differential operator composed of products of covariant derivatives and $\Gamma$ being any Dirac structure. Consider the following set of nonrenormalizable terms that contribute to the $ff\gamma$ vertex at the first order in $b_\nu$:
\begin{eqnarray}
{\cal L}^b_{\rm eff}&=&\frac{\alpha^{(5)}_1}{m_f}\,b^\mu\bar{\psi}\sigma_{\mu\nu}\gamma^5D^\nu\psi+\frac{\alpha^{(6)}_1}{m_f^2}\,b^\mu\bar{\psi}\gamma^\nu\gamma^5\stackrel{\Longleftrightarrow}{D_{\mu\nu}}\psi+\frac{\alpha^{(6)}_2}{m_f^2}\,b^\mu\bar{\psi}\gamma_\mu\gamma^5D^2\psi
\nonumber \\ &&
+\frac{\alpha^{(7)}_1}{m_f^3}\,ib^\mu\bar{\psi}\sigma_{\nu\lambda}\gamma^5\psi\partial_\mu F^{\nu\lambda}
+\frac{\alpha^{(7)}_2}{m_f^3}\,ib^\mu\bar{\psi}\sigma_{\mu\nu}\gamma^5\psi\partial_\lambda F^{\nu\lambda},
\label{stfobel}
\end{eqnarray}
with $D_{\mu\nu\ldots\rho}\equiv D_\mu D_\nu\ldots D_{\rho}$ and $D^2=D^\nu D_\nu$. The dimensionless coefficients $\alpha^{(i)}_j$ have two labels: the number within parentheses indicates the mass dimension of the corresponding nonrenormalizable operator; as it occurs with the second and third terms of ${\cal L}^b_{\rm eff}$, and with the fourth and fifth terms as well, the same mass dimension may be shared by more than one nonrenormalizable operators, which we label by means of the subscript in the $\alpha^{(i)}_j$ coefficients.

Powers of the fermionic mass correct units of the terms so that the total mass dimension of each one is four. The mass dimensions of these nonrenormalizable operators range from five to seven. After extracting from this Lagrangian all tree--level contributions to the $ff\gamma$ vertex, we find the vertex function
\begin{eqnarray}
\Gamma^{{\rm eff},b}_\mu&=&-\frac{\alpha^{(5)}_1}{m_f}\,i\sigma_{\mu\nu}b^\nu\gamma^5-\frac{\alpha^{(6)}_1}{m_f^2}\,b\cdot(p_1+p_2)\,\gamma_\mu\gamma^5-\frac{\alpha^{(6)}_2}{m_f^2}(p_{1\mu}+p_{2\mu})\slashed{b}\gamma^5
\nonumber \\ &&
+\frac{\alpha^{(7)}_1}{m_f^3}\,\frac{2i}{e}\,(b\cdot q)\,\sigma_{\mu\nu}q^\nu\gamma^5
+\frac{\alpha^{(7)}_2}{m_f^3}\,\frac{1}{e}\bigg[ b\cdot(p_1+p_2)\gamma^5\,q_\mu-i\,q^2\,\sigma_{\mu\nu}b^\nu\gamma^5\bigg],
\label{fobel}
\end{eqnarray}
which has all the Lorentz structures that are present in our result for the one--loop $ff\gamma$ term at the first order in $b_\nu$, Eq.~(\ref{order1r1}). As gauge invariance is an assumption imposed on ${\cal L}^b_{\rm eff}$ since the beginning, these tree--level contributions are gauge invariant. It is a straightforward task to write $\Gamma^b_\mu$ as Eq.~(\ref{fobel}), which proves that our one--loop result is gauge invariant.

The third term of $\Gamma_\mu^{{\rm eff},b}$, which is similar to the one associated to the electric dipole form factor, is generated by the third term of the effective Lagrangian ${\cal L}^b_{\rm eff}$. This operator is not related to the Lorentz and gauge invariant interaction
\begin{equation}
iF^{\mu\nu}\,\bar{\psi}\sigma_{\mu\nu}\gamma^5\psi,
\end{equation}
which conceives~\cite{ffapar} the electric dipole structure. One of the main differences among these operators is an extra derivative in the Lorentz--violating term, which produces, at the level of vertex function, an extra imaginary factor $i$. A similar reasoning can be utilized to see that the second term of $\Gamma_\mu^{{\rm eff},b}$ is not a contribution to the Lorentz--invariant axial current. 

All the terms of the effective Lagrangian ${\cal L}^b_{\rm eff}$ are $CPT$--odd and, indeed, they all share the same transformation properties with respect to the discrete transformations $C$, $P$, and $T$. Such properties of transformation are summarized in the first two rows of Table~\ref{cptpr}. We represent any nonrenormalizable operator at the first order in $b_\nu$ by $b_\mu{\cal O}^\mu(\psi,A_\rho)$, where we have explicitly indicated the Lorentz contraction of the four--vector $b_\nu$ with ${\cal O}^\nu$, which represents the rest of the nonrenormalizable structure. Any of such operators can then be split into two parts as
\begin{equation}
b_\mu{\cal O}^\mu(\psi, A_\rho)=b_0{\cal O}^0(\psi, A_\rho)+b_k{\cal O}^k(\psi, A_\rho),
\end{equation}
so that one term involves the timelike component $b_0$ and the other contains the spacelike ones, $b_k$. In Table~\ref{cptpr}, we have used this notation and provided the transformation properties under $C$, $P$, $T$, $CP$, and $CPT$ of $b_0{\cal O}^0$ and $b_k{\cal O}^k$ separately. The part carrying the $b_0$ timelike component in any nonrenormalizable term of ${\cal L}^b_{\rm eff}$ is $C$--even, $P$--odd, $T$--even, and, consequently, $CP$--odd, while the part involving the $b_k$ spacelike components is $C$--even, $P$-even, and $T$--odd, so that it is $CP$--even. It is worth emphasizing that these properties of transformation coincide with those~\cite{rcsme5} characterizing the renormalizable $CPT$--violating term of Eq.~(\ref{rcptvb}). In other words, the renormalizable $CPT$--violating axial coupling of $b_\nu$ to fermions has certain properties of transformation under $C$, $P$, and $T$, and such properties are preserved by the one--loop contributions to the $ff\gamma$ vertex produced by this Lorentz--violating interaction at the first order in $b_\nu$. More nonrenormalizable terms involving axial couplings of fermions to $b_\nu$ at the first order can be constructed, but they do not produce Lorentz structures of the one--loop $ff\gamma$ that we derived. The reason is that such nonrenormalizable terms have different properties with respect to the discrete transformations.

\begin{table}[ht]
\centering
\begin{tabular}{| c | c | c | c | c | c |}
\hline
 & $C$ & $P$ & $T$ & $CP$ & $CPT$ 
\\ \hline
$b_0{\cal O}^0$ & + & \textminus & + & \textminus & \textminus
\\[0.2cm]
$b_k{\cal O}^k$ & + & + & \textminus & + & \textminus
\\[0.1cm]\hline
$b^2{\cal O}$ & + & + & + & + &
+
\\[0.2cm]
$b_0 b_0{\cal O}^{00}$ & + & + & + & + & +
\\[0.2cm]
$b_k b_n{\cal O}^{kn}$ & + & + & + & + & +
\\[0.2cm]
$b_0 b_k\left({\cal O}^{0k}+{\cal O}^{k0}\right)$ & + & \textminus & \textminus & \textminus & +
\\ \hline
\end{tabular}
\caption{\label{cptpr} Properties of transformation under $C$, $P$, $T$, $CP$, and $CPT$ of nonrenormalizable operators.}
\end{table}

We introduce the following set of nonrenormalizable operators, with mass dimensions ranging from six to nine, that contribute to the $ff\gamma$ vetex at tree level:
\begin{eqnarray}
{\cal L}_{\rm eff}^{b^2}&=&\frac{\beta^{(6)}_1}{m_f^2}\,ib^2\,\bar{\psi}\gamma_\mu D^\mu\psi+\frac{\beta^{(6)}_2}{m_f^2}\,ib^\nu b^\rho\,\bar{\psi}\gamma_\nu D_\rho\psi+\frac{\beta^{(7)}_1}{m_f^3}\,b^2F^{\mu\nu}\,\bar{\psi}\sigma_{\mu\nu}\psi
\nonumber \\ &&
+\frac{\beta^{(7)}_2}{m_f^3}\,b^\rho b^\nu\,\bar{\psi}D_{\rho\nu}\psi
+\frac{\beta^{(7)}_3}{m_f^3}\,b^\rho b_\lambda F_{\rho\nu}\,\bar{\psi}\sigma^{\nu\lambda}\psi
+\frac{\beta^{(8)}_1}{m_f^4}\,b^2\,\bar{\psi}\gamma_\mu\psi \partial_\nu F^{\mu\nu}
\nonumber \\&&
+\frac{\beta^{(8)}_2}{m_f^4}\,b_\nu b_\lambda\,\bar{\psi}\gamma_\rho\psi\,\partial^\nu F^{\rho\lambda}+\frac{\beta^{(8)}_3}{m_f^4}\,ib_\rho b_\lambda\,\bar{\psi}\gamma_\nu D^{\rho\nu\lambda}\psi+\frac{\beta^{(8)}_4}{m_f^4}\,b^\rho b^\mu\,\bar{\psi}\gamma_\mu\psi\,\partial^\nu F_{\rho\nu}
\nonumber \\ &&
+\frac{\beta^{(8)}_5}{m_f^4}\,ib^\rho b^\lambda\,\bar{\psi}\gamma_\lambda\stackrel{\longleftrightarrow}{D_{\rho\nu}\hspace{0.00001cm}^\nu}\psi+\frac{\beta^{(9)}_1}{m_f^5}\,b^\rho b^\lambda F^{\mu\nu}\,(\overline{D_\rho\psi})\sigma_{\mu\nu}D_\lambda\psi+\frac{\beta^{(9)}_2}{m_f^5}\,b^\rho b^\lambda\,\bar{\psi}D^\nu\hspace{0.00001cm}_{\rho\lambda\nu}\psi
\nonumber \\ &&
+\frac{\beta^{(9)}_3}{m_f^5}\,ib^\rho b^\lambda F_{\rho\nu}\left[ (\overline{D^\nu\psi})D_\lambda\psi-(\overline{D_\lambda\psi})D^\nu\psi \right].
\end{eqnarray}
This Lagrangian parametrizes effects from a high--energy description through the dimensionless parameters $\beta^{(i)}_j$. The labels in these coefficients follow the same notation of the linear case, for which we employed the $\alpha^{(i)}_j$. The nonrenormalizable terms contained in ${\cal L}^{b^2}_{\rm eff}$ carry contributions from the background vector $b_\nu$ at the second order. We have included a larger number of terms than in the case of the $b_\nu$ first--order contributions in order to generate all the Lorentz structures of the one--loop expression $\Gamma^{b^2}_\mu$, Eq.~(\ref{sovf}). This effective Lagrangian produces the vertex function
\begin{eqnarray}
\label{effsovf}
\Gamma_\mu^{{\rm eff},b^2}&=&-\frac{\beta^{(6)}_1}{m_f^2}\,b^2\gamma_\mu-\frac{\beta^{(6)}_2}{m_f^2}\,b_\mu\slashed{b}-\frac{\beta^{(7)}_1}{m_f^3}\,\frac{2}{e}\,b^2\,i\sigma_{\mu\nu}q^\nu+\frac{\beta^{(7)}_2}{m_f^3}\,b_\mu b\cdot(p_1+p_2)
\nonumber \\ \nonumber &&
+\frac{\beta^{(7)}_3}{m_f^3}\frac{1}{e}\left[ i\,(b\cdot q)\,\sigma_{\mu\nu}b^\nu+2m_f\,b_\mu\slashed{b}-b_\mu b\cdot(p_1+p_2) \right]
+\frac{\beta^{(8)}_1}{m_f^4}\frac{1}{e}\,b^2q^2\gamma_\mu
\\ \nonumber &&
+\frac{\beta^{(8)}_2}{m_f^4}\,\frac{1}{e}\left[ \left( (b\cdot p_1)^2+(b\cdot p_2)^2 \right)\,\gamma_\mu-2\,(b\cdot p_1)(b\cdot p_2)\,\gamma_\mu \right]
\nonumber \\&&
+\frac{\beta^{(8)}_3}{m_f^4}\,\Big[ m_e\,b_\mu\,b\cdot(p_1+p_2)+\gamma_\mu(b\cdot p_1)(b\cdot p_2) \Big]
+\frac{\beta^{(8)}_4}{m_f^4}\,\frac{1}{e}\left[ -(b\cdot q)\,\slashed{b}q_\mu+b_\mu\slashed{b}\,q^2 \right]
\nonumber \\ &&
+\frac{\beta^{(8)}_5}{m_f^4}\,\bigg[ 2m_f^2\,b_\mu\slashed{b}-(b\cdot q)\slashed{b}q_\mu+2\slashed{b}\Big( p_{1\mu}(b\cdot p_1)+p_{2\mu}(b\cdot p_2) \Big) \bigg]
\nonumber \\ &&
-\frac{\beta^{(9)}_1}{m_f^5}\,\frac{2i}{e}\,(b\cdot p_1)(b\cdot p_2)\,\sigma_{\mu\nu}q^\nu
\nonumber \\&&
+\frac{\beta^{(9)}_2}{m_f^5}\,\left[ b_\mu\, b\cdot(p_1+p_2)\left( m_f^2-\frac{q^2}{2} \right)+p_{1\mu}\,(b\cdot p_1)^2+p_{2\mu}\,(b\cdot p_2)^2 \right]
\nonumber \\ &&
+\frac{\beta^{(9)}_3}{m_f^5}\frac{1}{e}\,\bigg[ \frac{q^2}{2}\,b_\mu\,b\cdot(p_1+p_2)-2m_f\,\gamma_\mu\,(b\cdot p_1)(b\cdot p_2)
\nonumber \\&&
+p_{1\mu}\,(b\cdot p_2)^2+p_{2\mu}\,(b\cdot p_1)^2
+i\,(b\cdot p_1)(b\cdot p_2)\,\sigma_{\mu\nu}q^\nu
\bigg]
\end{eqnarray}
The one--loop contributions to the $ff\gamma$ vertex at the second order in $b_\nu$ that we calculated, Eq.~(\ref{sovf}), can be expressed in the same form as $\Gamma^{{\rm eff},b^2}_\mu$, which, by construction, is gauge invariant. Therefore, the $b_\nu$--second--order contributions $\Gamma_\mu^{b^2}$ of our one--loop result are gauge invariant as well.

The behavior under $C$, $P$, $T$, $CP$, and $CPT$ of the set of nonrenormalizable operators involving $b_\nu$--second--order contributions is summarized in Table~\ref{cptpr}, from the third to the sixth rows. Similarly to what we did in the case of the $b_\nu$--first--order contributions, we generically represent any nonrenormalizable operator at the second order in $b_\nu$ by $b_\mu b_\nu{\cal O}^{\mu\nu}(\psi,A_\rho)$, where the product $b_\mu b_\nu$ is Lorentz--contracted with ${\cal O}^{\mu\nu}$ to constitute the whole nonrenormalizable operator, which can be divided as a sum of three terms:
\begin{eqnarray}
b_\mu b_\nu{\cal O}^{\mu\nu}(\psi,A_\rho)&=&b_0b_0{\cal O}^{00}(\psi,A_\rho)+b_kb_n{\cal O}^{kn}(\psi,A_\rho)
\nonumber \\&&
+b_0b_k\left[ {\cal O}^{0k}(\psi,A_\rho)+{\cal O}^{k0}(\psi,A_\rho) \right],
\end{eqnarray}
each one containing a product of timelike components $b^0b^0$, or a product of spacelike parameters $b^kb^n$, or a product $b^0b^k$ of a timelike component with a spacelike component. We have utilized this notation in Table~\ref{cptpr}, where we have included the transformation properties of the $b_0b_0{\cal O}^{00}$, $b_kb_n{\cal O}^{kn}$ and $b_0b_k\left( {\cal O}^{0k}+{\cal O}^{k0} \right)$ terms. Note that $b^2=g_{\mu\nu}b^\mu b^\nu$, so that the first, third, and sixth terms of ${\cal L}^{b^2}_{\rm eff}$ contain parts proportional to $b^0b^0$ and $b^kb^n$, which consistently possess the correct transformation properties. Nevertheless, we show the transformation properties of these terms separately in the table. Contrastingly to the $b_\nu$--first--order effective Lagrangian, ${\cal L}^b_{\rm eff}$, all the nonrenormalizable terms in ${\cal L}^{b^2}_{\rm eff}$ are $CPT$--even, which means that the contributions at the second order in $b_\nu$ to the $ff\gamma$ vertex are invariant under $CPT$. The first and third terms of ${\cal L}^{b^2}_{\rm eff}$ are even with respect to the discrete transformations $C$, $P$, and $T$, so that these operators are $CP$--even. The remaining nonrenormalizable operators have common transformation properties, since, in all cases, the parts proportional to $b^0b^0$ or $b^kb^n$ are even under $C$, $P$, $T$, and $CP$, whereas those involving products $b^0b^k$ are $C$--even, $P$--odd, $T$--odd, and $CP$--odd. As before, notice that more independent gauge invariant nonrenormalizable structures can be realized, but they will not generate terms that are present in our one--loop result, $\Gamma_\mu^{b^2}$, as their properties with respect to the discrete transformations are different.

As it can be appreciated in Eq.~(\ref{sovf}), the one--loop contribution to the $ff\gamma$ vertex at the second order in $b_\nu$ produces\footnote{Contributions to the electric dipole form factor were not generated. However, they may be produced at the two--loop level.} a vector current and the Lorentz structures that define the magnetic dipole form factor and the anapole from factor, being all of them proportional to $b^2$. The fact that these terms entail genuine contributions to such quantities relies on the first, third and sixth nonrenormalizable operators of the effective Lagrangian ${\cal L}^{b^2}_{\rm eff}$, which generate such structures at tree level. As $b^2$ is a scalar quantity, these terms match the gauge and Lorentz invariant Lagrangian terms
\begin{eqnarray}
\label{vcLs}
&&i\bar{\psi}\gamma_\mu D^\mu\psi,
\\ \nonumber \\
&&\label{mdLs}
F^{\mu\nu}\bar{\psi}\sigma_{\mu\nu}\psi,
\\Ê\nonumber \\
&&\label{aLs}
\bar{\psi}\gamma_\mu\psi\partial_\nu F^{\mu\nu},
\end{eqnarray}
which give rise to the well--known form factors~\cite{ffapar,NRST}. There exist other terms in the one--loop expression $\Gamma^{b^2}_\mu$ that seem to contribute to these quantities, but, as in the case of the $b_\nu$--first--order result, their nonrenormalizable generating operators are not related to the Lagrangian terms (\ref{vcLs}), (\ref{mdLs}), and (\ref{aLs}), so they are not really contributions. In the next section, we discuss phenomenological consequences of the $b_\nu$ second--order contributions to the Lorentz--invariant parametrization of the $ff\gamma$ vertex and set a bound on $CPT$ violation through the scalar $b^2$.

\section{Bounds on $CPT$ violation}
\label{dscn}
The discussion in the last paragraph of the previous section ensures that the $CPT$--violating axial coupling of $b_\nu$ to fermions, added to the Lorentz--invariant Dirac Lagrangian, actually produces contributions to low--energy quantities through the one--loop $ff\gamma$ vertex. A one--loop calculation of the effects from the isotropic--Lorentz--violating photon sector to this vertex has been already carried out~\cite{rcsme7}. In such context, a contribution to the AMM emerges and a contribution to the vector current is also produced~\cite{MNTT}, both at the first order in Lorentz violation. This is different with respect to what we have found in the present paper, since we had to go further and calculate the second--order Lorentz--violating contributions to produce the same Lorentz structures. The vector current arisen from our calculation involves an ultraviolet--finite contribution to the corresponding form factor. However, some renormalization scheme should conveniently absorb it. The anapole moment, on the other hand is a gauge dependent quantity~\cite{NRST}. For this reasons, we concentrate on the new--physics contributions to the AMM.

The calculation of the SM contributions to the AMM of the electron and the muon has been performed with great precision~\cite{AHKNe,Krev,AHKNm} and accurate experiments have yielded~\cite{expamm1,expamm2,expamm3,expamm4} precise measurements of these quantities as well. Recently, in Refs.~\cite{AHKNe,AHKNm}, the disagreements between these values, for the cases of the electron and the muon, were reported to be $\Delta a_e=a_e^{\rm EXP}-a_e^{\rm SM}=-1.06\times10^{-12}$ and $\Delta a_\mu=a_\mu^{\rm EXP}-a_\mu^{\rm SM}=2.49\times10^{-9}$. Though these discrepancies might be reduced by further studies, they can be used to constrain new physics. As the most general version of the minimal SME dictates~\cite{CK1}, there is a Lorentz--violating background vector $b_\nu$ for each fermion. With minor amendments concerning electric charges, our result could then be used to set a bound on Lorentz violation associated to any charged lepton and to any quark if differences such as $\Delta a_e$ and $\Delta a_\mu$ were available for all these particles. Due to the lack of such data, we restrict our analysis to the cases of the electron and the muon.

According to the Lorentz--invariant parametrization of the $ff\gamma$ vertex~\cite{ffapar} and to Eq.~(\ref{sovf}), the contribution to the magnetic dipole form factor, $a_f(q^2)$, that arises from the $CPT$--violating renormalizable axial coupling of $b_\nu$ to fermions is given by
\begin{equation}
a_f(q^2)=\frac{2b^2}{m_f^2}\,F^{b^2}_3(q^2),
\end{equation}
with $F^{b^2}_3(q^2)$ explicitly shown in Eq.~(\ref{piamm}). The AMM is defined by the setting $q^2=0$ in the magnetic dipole form factor $a_f(q^2)$. A nice feature of the contribution to the magnetic dipole form factor that we derived is that it involves effects from the SME that are parametrized by a Lorentz scalar, $b^2$, which means that the corresponding contribution to AMM does not depend on a particular reference frame. Imposing the conditions $|a_e(0)|<|\Delta a_e|$ and $|a_\mu(0)|<|\Delta a_\mu|$, and omitting infrared--divergent terms, we obtain bounds on the magnitudes of the $CPT$--violating background four--vectors $b_\nu$ associated with the electron and the muon:
\begin{eqnarray}
|b^{e^-}|&<&8.06203\times10^{-9},
\\ \nonumber \\
|b^{\mu^-}|&<&8.07933\times10^{-5}.
\end{eqnarray}

The $CPT$ violating background vector $b_\nu$ has been stringently constrained by different means that include Hg/Cs comparison~\cite{expsme1,hgcsexpsme2}, Penning trap~\cite{expsme2}, torsion pendulum~\cite{expsme5,expsme6,expsme8}, K/He magnetometer~\cite{mgnmtexpsme}, $g_\mu-2$ data~\cite{expsme3,gm2expsme2,expsme7}, and munonium spectroscopy~\cite{expsme4}. The bounds on the components of this Lorentz--violating four--vector range from $10^{-27}$ to $10^{-31}$ for the electron, and from $10^{-22}$ to $10^{-24}$ for the muon. Clearly, these constraints are much more restrictive than the bounds derived in the present paper. This is partly due to the fact that the new--physics contributions arose only since the second order in the $b_\nu$ vector, and not before. Note, on the other hand, that the sizes of the discrepancies $\Delta a_f$, though small, are not enough restrictive to provide good bounds on Lorentz violation. However, we would like to emphasize the way in which this result was achieved, as well as the independence of our bounds with respect to particular reference frames. 

\section{Conclusions}
\label{conc}
In the present paper we calculated and analyzed contributions from the QED sector of the minimal SME to the $ff\gamma$ vertex at the one--loop level. Our study comprehended the one--loop calculation, the construction of a set of nonrenormalizable terms to prove gauge invariance of our results, and bounds on Lorentz violation from contributions to AMM in this vertex.
We started from the renormalizable $CPT$--violating effects parametrized by the background vector $b_\nu$, which couples axially to fermions in the Dirac Lagrangian. Such coupling yields a modified fermionic propagator that we utilized to derive an expression for the contributions from this vector background to the one--loop $ff\gamma$ vertex. We took the external photon off shell and the external fermions on shell, and then calculated all new--physics contributions to the $ff\gamma$ vertex at the first and second orders in $b_\nu$. It occurred that simple Ward identities were not obtained by contracting the momentum of the external photon with the resulting $ff\gamma$ amplitude. So, to prove gauge invariance, we constructed a set of gauge invariant nonrenormalizable operators contributing, at tree level, to the $ff\gamma$ vertex and generating all Lorentz structures arisen in the one--loop calculation. The idea was to keep in mind that gauge invariance is one of the building blocks of these nonrenormalizable invariants, so that their tree--level contributions are governed by gauge symmetry. Consequently, by generating in this manner all Lorentz structures present in the one--loop result, one can conclude that such expression is gauge invariant. The $b_\nu$--first--order contributions were parametrized by five $CPT$--violating nonrenormalizable terms of mass dimensions ranging from five to seven. These operators possess the same transformation properties under $C$, $P$, and $T$ that characterize the renormalizable axial coupling introduced in the Dirac Lagrangian. SME contributions to low--energy physics emerge in the one--loop $ff\gamma$ vertex since the second order in $b_\nu$. We found that $b_\nu$--second--order contributions are produced by a set of nonrenormalizable operators of mass dimensions six to nine that, contrastingly to the first--order case, are $CPT$--even. We constructed thirteen independent nonrenormalizable terms, at the second order in $b_\nu$, that together produce all Lorentz structures contained in the one--loop expression. One of these $b_\nu$--second--order terms coincide with the nonrenormalizable operator that produce, at the level of vertex function, the Lorentz structure defining the magnetic dipole form factor, which generates a contribution to AMM when taking the condition $q^2=0$ on the momentum of the external photon. Once gauge invariance was proven, our discussion centered on the possibility of constraining Lorentz violation by means of our result. The contributions to AMM arisen from this one--loop calculation are parametrized by the Lorentz scalar $b^2$, which is independent of reference frames and which we used to set bounds on $CPT$--violating $b_\nu$ vectors. We set bounds on the magnitudes of the background vectors $b^{e^-}$ and $b^{\mu^-}$, respectively corresponding to the electron and the muon, by comparing the corresponding contributions to AMM with the differences $\Delta a_f$ between the SM prediction and the experimentally measured values of the AMM. We determined that such discrepancies are not tiny enough to set strong bounds on the $b_\nu$ vectors associated to the electron and the muon, since we found constraints that are much less restrictive than the most stringent bounds currently reported in the literature.

\section*{Acknowledgments}

We acknowledge financial support from CONACYT and SNI (M\' exico). 

\appendix

\section{List of parametric integrals (first order contributions)}
\label{fopi}
The dimensionless factors involved in the $CPT$--violating contributions at the first order in $b_\mu$, Eq.~(\ref{order1r1}), are defined as
\begin{eqnarray}
F^b_1(q^2)&=&\frac{\alpha\, m_f^2}{\pi}\int_0^1dx\int_0^{1-x}dy\,\Bigg\{ \frac{x(3x-3y-1)}{m_f^2(x+y)^2-q^2xy}
\nonumber \\  &&
+\frac{1}{\left( m_f^2(x+y)^2-q^2xy \right)^2}\Big[ 2m_f^2\left( 4y-(x+y)(x^3-(x+1)y^2+3(x+2)y) \right)
\nonumber \\&&
+2q^2y(x^3-x^2y+x^2+x+2y-2) \Big]
\Bigg\},
\\ \nonumber \\
F^b_2(q^2)&=&\frac{\alpha\,m_f^2}{\pi}\int_0^1dx\int_0^{1-x}dy\,\Bigg\{ -\frac{(1-x)(2x-y)}{m_f^2(x+y)^2-q^2xy}
\nonumber \\ &&
\frac{1}{\left( m_f^2(x+y)^2-q^2xy \right)^2}
\Big[2m_f^2\left( (x+y)(x^2+(x-3)y^2+y^3+4y)-2y \right)
\nonumber \\&&
-2q^2(x-1)(y-1)^2y\Big] \Bigg\},
\\ \nonumber \\
F^b_3(q^2)&=&\frac{\alpha\,m_f^2}{\pi}\int_0^1dx\int_0^{1-x}dy\,\Bigg\{
\frac{x(x-2y)}{m_f^2(x+y)^2-q^2xy}
\nonumber \\ &&
+\frac{2m_f^2\,x(x+y)(x(y-1)+(y-3)y)+2q^2\,xy(-xy+x+y-1)}{\left( m_f^2(x+y)^2-q^2xy \right)^2}
\Bigg\},
\\ \nonumber \\
F_4^b(q^2)&=&\frac{2\alpha\,m_f^4}{\pi}\int_0^1dx\int_0^{1-x}dy\,\frac{(x-y)(x+y-1)(x^2-xy+y)}{\left( m_f^2(x+y)^2-q^2xy \right)^2},
\\ \nonumber \\
F_5^b(q^2)&=&-\frac{2\alpha\,m_f^4}{\pi}\int_0^1dx\int_0^{1-x}dy\,\frac{(x+y-1)(x+y)(x^2-xy+y)}{\left( m_f^2(x+y)^2-q^2xy \right)^2},
\end{eqnarray}
where $\alpha$ is the fine--structure constant.

\section{List of parametric integrals (second order contributions)}
\label{sopi}
The dimensionless factors generated by the $b_\mu$-second--order contributions, Eq.~(\ref{sovf}), are
\begin{eqnarray}
F_1^{b^2}(q^2)&=&\frac{\alpha m_f^2}{2\pi}\int_0^1dx\int_0^{1-x}dy\,\Bigg\{ \frac{4x y}{m_f^2 (x+y)^2-q^2 x y}
\nonumber \\ \nonumber &&
+\frac{2m_f^2xy\Big( -(x+y-1)
   (3 x+3 y-5)+4(x (x+y-6)-y+3) \Big)}{\left(m_f^2 (x+y)^2-q^2 x y\right){}^2}
\nonumber \\ \nonumber &&
+\frac{6m_f^4xy}{\left(m_f^2 (x+y)^2-q^2 x y\right){}^3}\Big[\Big(x^4+4 x^3 (y-1)+2 x^2
   (3 (y-2) y+2)
\nonumber \\&&   
+4 x ((y-2) (y-1) y+2)
+y \left(y
(y-2)^2+8\right)-8\Big)
\nonumber \\ &&
+8 \left(2 x^3+x^2 (4 y-3)+2 x(y-1)^2-y^2\right)\Big]
\Bigg\},
\end{eqnarray}
\begin{eqnarray}
F^{b^2}_2(q^2)&=&\frac{\alpha m_f^2}{\pi}\int_0^1dx\int_0^{1-x}dy\,\Bigg\{
-\frac{2x y}{m_f^2 (x+y)^2-q^2 x y}
\nonumber \\ \nonumber &&
+\frac{x y \left(2 m_f^2 \left(2 x^2+4 x (y-1)+2 (y-2)
   y+1\right)+q^2 (x (2-4 y)+2 y-1)\right)}{\left(m_f^2 (x+y)^2-q^2 x y\right){}^2}
\\ \nonumber &&
+\frac{6xy}{\left(m_f^2 (x+y)^2-q^2 x y\right){}^3}\Big[ q^2 m_f^2 (x+y-2) ((x+y) (x (2
   y-1)-y)+2)
\\ \nonumber &&
   -m_f^4 \Big(x^4+4 x^3 (y-1)+2 x^2 (3(y-2) y+2)+4 x ((y-2) (y-1) y+2)
\\ &&
+y \left(y(y-2)^2+8\right)-8\Big)+(q^2)^4 x y (-xy+x+y-1) \Big]
\Bigg\},
\end{eqnarray}
\begin{eqnarray}
F^{b^2}_3(q^2)&=&\frac{2\alpha m_f^4}{\pi}\int_0^1dx\int_0^{1-x}dy\,
\Bigg\{
-\frac{x y (x (x+y-6)-y+3)}{\left(m_f^2 (x+y)^2-q^2 x y\right){}^2}
\nonumber \\ &&
+\frac{6x y }{\left(m_f^2 (x+y)^2-q^2 x y\right){}^3}\Big(m_f^2 \left(-2 x^3+x^2 (3-4 y)-2 x
   (y-1)^2+y^2\right)
\nonumber \\&&
   +q^2 (x-1) x (2 y-1)\Big)
\Bigg\},
\label{piamm}
\end{eqnarray}
\begin{eqnarray}
F^{b^2}_4(q^2)&=&\frac{2\alpha m_f^4}{\pi}\int_0^1dx\int_0^{1-x}dy\Bigg\{
-\frac{x y (x+y-1) (x+y)}{\left(m_f^2 (x+y)^2-q^2 x y\right){}^2}
\nonumber \\ \nonumber &&
+\frac{3x y (x+y-1) }{\left(m_f^2 (x+y)^2-q^2 x y\right){}^3}\Big[m_f^2 \big(x^3+x^2 (3 y-1)+x (y
   (3 y-2)-2)
\nonumber \\&&
+(y-2) y (y+1)+4\big)
+q^2 \left(-x
   \left(x (y-1)+y^2\right)+y-2\right)\Big]
\Bigg\},
\end{eqnarray}
\begin{eqnarray}
F^{b^2}_5(q^2)&=&\frac{2\alpha m_f^4}{\pi}\int_0^1dx\int_0^{1-x}dy\,\Bigg\{
\frac{x y \left(-x^2+x+y^2+y-1\right)}{\left(m_f^2 (x+y)^2-q^2 x y\right){}^2}
\nonumber \\ \nonumber &&
+\frac{3x y}{\left(m_f^2 (x+y)^2-q^2 x y\right){}^3} \Big[m_f^2 \Big(x^4+2 x^3 (y-1)+x^2 (1-4 y)
\nonumber \\&&
-2x \left(y \left(y^2+y-3\right)+3\right)
-y\left(y^3-5 y+6\right)+4\Big)
\nonumber \\&&
-q^2 \left(x^3(y-1)+x^2-x \left(y\left(y^2+y-1\right)+2\right)+y^2-3y+2\right)\Big]
\Bigg\},
\end{eqnarray}
\begin{eqnarray}
F^{b^2}_6(q^2)&=&\frac{\alpha m_f^4}{2\pi}\int_0^1dx\int_0^{1-x}dy\,\Bigg\{
\frac{xy(x (6 y-4)-4 y+3)}{\left(m_f^2 (x+y)^2-q^2 x y\right){}^2}
\nonumber \\&&
+\frac{6xy}{\left(m_f^2 (x+y)^2-q^2 x y\right){}^3}\Bigg[q^2 (x-1) x (y-1)y
\nonumber \\&&
+m_f^2\Big(8x (-2 x y+x+2y-1)
\nonumber \\ &&
-(x+y-2) ((x+y) (x (2y-1)-y)+2)\Big)\Bigg]
\Bigg\}
\end{eqnarray}
\begin{eqnarray}
F^{b^2}_7(q^2)&=&\frac{2\alpha m_f^4}{\pi}\int_0^1dx\int_0^{1-x}dy\,\Bigg\{
\frac{x y(x-1)^2}{\left(m_f^2 (x+y)^2-q^2 x y\right){}^2}
\nonumber \\&&
-\frac{6m_f^2\,x y \left(x^3-x^2-x (y-2) y+y^2\right)}{\left(m_f^2 (x+y)^2-q^2 x y\right){}^3}
\Bigg\},
\end{eqnarray}
\begin{eqnarray}
F^{b^2}_8(q^2)&=&\frac{2\alpha m_f^4}{\pi}\int_0^1dx\int_0^{1-x}dy\,\Bigg\{
\frac{x y (2 x y-1)}{\left(m_f^2 (x+y)^2-q^2 x y\right){}^2}
\nonumber \\ \nonumber &&
-\frac{24m_f^2\,x y \left(x^2+(x-1) y\right) (x (2 y-1)-y+1)}{\left(m_f^2 (x+y)^2-q^2 x y\right){}^3}
\nonumber \\&&
+\frac{6x y}{\left(m_f^2 (x+y)^2-q^2 x y\right){}^3}\Big[ m_f^2 \big(x^2 (3-4 y)-2 x (y-2) (2y-1)
\nonumber \\&&
+y (3 y-4)+2\big)
+q^2 (x-1) (y-1)(x+y-1)\Big]
\Bigg\},
\end{eqnarray}
\begin{eqnarray}
F^{b^2}_9(q^2)&=&\frac{2\alpha m_f^4}{\pi}\int_0^1dx\int_0^{1-x}dy\,\Bigg\{
\frac{x^2 y (2 y-1)}{\left(m_f^2 (x+y)^2-q^2 x y\right){}^2}
\nonumber \\ &&
+\frac{6x^2 y }{\left(m_f^2 (x+y)^2-q^2 x y\right){}^3}\Big[q^2 (x-1) (y-1) y
\nonumber \\&&
-m_f^2 \left(x^2
   (y-1)+2 x (y-1)^2+(y-2) (y-1) y-2\right)\Big]
\Bigg\},
\end{eqnarray}
\begin{eqnarray}
F^{b^2}_{10}(q^2)&=&\frac{2\alpha m_f^4}{\pi}\int_0^1dx\int_0^{1-x}dy\,\Bigg\{
-\frac{x y (2 x (x+y-2)+1)}{\left(m_f^2 (x+y)^2-q^2 x y\right){}^2}
\nonumber \\ \nonumber &&
+\frac{6x y}{\left(m_f^2 (x+y)^2-q^2 x y\right){}^3}\Big[m_f^2 \Big(x^4+x^3 (3 y-5)+x^2 (y-1) (3y-7)
\\ &&
+x (y-2)^2 (y-1)+y^2\Big)
-q^2 (x-1) x(y-1) (x+y-1)\Big]
\Bigg\},
\end{eqnarray}
\begin{equation}
F^{b^2}_{11}(q^2)=\frac{24\alpha m_f^6}{\pi}\int_0^1dx\int_0^{1-x}dy\,\frac{x y \left(x^2+(x-1) y\right) (x (2 y-1)-y+1)}{\left(m_f^2 (x+y)^2-q^2 x y\right){}^3},
\end{equation}
\begin{eqnarray}
F^{b^2}_{12}(q^2)&=&\frac{24\alpha m_f^6}{\pi}\int_0^1dx\int_0^{1-x}dy\,\frac{(1-x) x^2 y \left(x^2+(x-1) y\right)}{\left(m_f^2 (x+y)^2-q^2 x y\right){}^3},
\end{eqnarray}
\begin{equation}
F^{b^2}_{13}(q^2)=\frac{24\alpha m_f^6}{\pi}\int_0^1dx\int_0^{1-x}dy\,\frac{x (1-y) y^2 \left(x^2+(x-1) y\right)}{\left(m_f^2 (x+y)^2-q^2 x y\right){}^3}.
\end{equation}

\end{document}